\documentclass[lettersize,journal]{IEEEtran}
\usepackage{cite,graphicx,amsmath,array,subfigure,url,booktabs,xcolor,multirow,color,balance,amsfonts,algorithmic,algorithm,textcomp,stfloats,verbatim}
\begin{document}

\title{Prototype of Secure Wire-Line Telephone}

\author{
	Lifeng Lin, \IEEEmembership{Graduate Student Member,~IEEE},
	Zijian Zhou, \IEEEmembership{Member,~IEEE},
	Peihe Jiang,
	Sanjun Liu,\\
	Lai Wei, \IEEEmembership{Member,~IEEE}, and
	Bingli Jiao, \IEEEmembership{Senior Member,~IEEE}
	\thanks{This work was jointly supported by the Science and Technology Project of Guangzhou under Grants 202206010118 and 2023B04J0011, and the National Natural Science Foundation of China under Grant 62171006.  The calculations were supported by the High-Performance Computing Platform of Peking University.  \emph{(Corresponding author: Zijian Zhou.)}}
	\thanks{Lifeng Lin, Zijian Zhou, Lai Wei, and Bingli Jiao are with the School of Electronics, Peking University, Beijing 100871, China (e-mail: \{linlifeng, zjzhou1008, future1997, jiaobl\}@pku.edu.cn).}
	\thanks{Peihe Jiang is with the College of Physics and Electronic Information, Yantai University, Yantai 264005, China (e-mail: jiangpeihe@ytu.edu.cn).}
	\thanks{Sanjun Liu is with the College of Intelligent Systems Science and Engineering, Hubei Minzu University, Enshi 445000, China (e-mail: liusanjunbox1@126.com).}
}
\maketitle

\begin{abstract}
This paper presents a secure wire-line telephone system that employs physical layer security (PLS) to protect against wiretapping.  The system generates artificial noise (AN) in both transmission directions and uses a telephone hybrid circuit to effectively suppress the AN for the purpose of secure communication.  Furthermore, we analyze the secrecy capacity of the system and evaluate its performance through theoretical analysis and practical experiments.  The results demonstrate that the proposed system can significantly enhance communication security while preserving the integrity of legitimate signals.  The results also validate that the proposed system is a robust and effective solution for securing wire-line telephone communications.
\end{abstract}

\section{Introduction}
\IEEEPARstart{T}{elephone} systems have been widely used for both voice and data transmission in a full-duplex manner for many years. Traditionally, the privacy of these communications has been safeguarded using cryptographic methods \cite{ref_Shannon_Bell_1949}, where cryptographic codes have effectively protected messages from wiretappers. However, these methods are becoming increasingly vulnerable to decryption due to the advancement of powerful computers. As computing power continues to grow, the time required to break cryptographic codes decreases, and this necessitates the exploration of alternative methods for securing communications.

Physical layer security (PLS) \cite{ref_Wyner_Bell_1975, ref_Poor_PNAS_2017} offers a fundamentally different approach to securing communications without relying on cryptography. In PLS, the security model involves two legitimate users, Alice and Bob, and a malicious eavesdropper, Eve. Absolute communication security is theoretically achievable if Eve's channel conditions, particularly her signal-to-noise ratio (SNR), are worse than those of Alice and Bob. Specifically, if Eve's SNR is lower than -1.59 dB, it becomes impossible for her to extract any meaningful information from the transmitted signals.

To achieve absolute security through PLS, researchers have proposed using artificial noise (AN) to degrade Eve's channel quality for wireless communications \cite{ref_Sobers_TWC_2017, ref_Li_TWC_2022, ref_Zhou_TVT_2018, ref_Zhou_TVT_2022, ref_Tang_CL_2023}.  By adding the AN to the communication channel, these methods increase Eve's interference level and prevent her from successfully intercepting the communication.  However, a significant limitation of this approach is the rapid attenuation of the AN over distance for wireless propagation.  If Eve is located close to signal sources and far from the AN sources, the AN may not sufficiently degrade her channel and it thus leaves the communication vulnerable to eavesdropping.

In contrast, wire-line telephone systems experience much lower signals or ANs attenuation over the distance \cite{ref_Liu_CL_2017}.  For example, the AN transmitted by Bob along a twisted differential line may attenuate by only a few decibels per hundred meters.  This low attenuation allows Bob to transmit AN that effectively conceals Alice's messages over distances of up to a thousand meters.  This characteristic makes wire-line telephone systems particularly well-suited for PLS techniques that use AN.

Motivated by these considerations, we propose a new secure wire-line telephone system that can generate the AN to protect data transmission.  Unlike the approach in \cite{ref_Liu_CL_2017}, our focus is to safeguard data security in bidirectional communication and to provide a practical prototype.  Inspired by the co-frequency co-time full-duplex (CCFD) technology and its potentials \cite{ref_Ding_arXiv1_2024, ref_Ding_arXiv2_2024, ref_Ding_arXiv3_2024, ref_Ding_arXiv4_2024, ref_Lin_arXiv1_2024, ref_Zhou_arXiv1_2024}, we propose an AN cancellation method based on a telephone hybrid circuit and provide an analytical expression for its performance.  Furthermore, extensive experimental results validate the AN cancellation capability of the telephone hybrid circuit and demonstrate the effectiveness of the proposed system in ensuring secure communications.

The remainder of the paper is organized as follows.  Section\;\ref{sec2} introduces the system model and analyzes the secrecy capacity based on information theory.  In Section\;\ref{sec3}, we propose the hardware implementation and the telephone hybrid circuit for the secure wire-line telephone system.  Experimental results are carried out in Section\;\ref{sec4} and the paper is concluded in Section\;\ref{sec5}.

\section{System Model} \label{sec2}
\begin{figure}
	\centering
	\subfigure[System model]{ \label{fig1a}
		\centering
		\includegraphics[width=0.9\columnwidth]{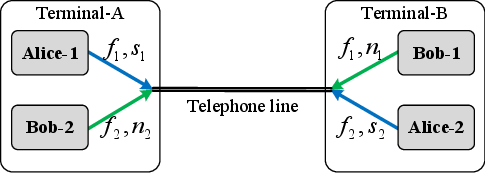}}
	\subfigure[Security model]{ \label{fig1b}
		\centering
		\includegraphics[width=0.9\columnwidth]{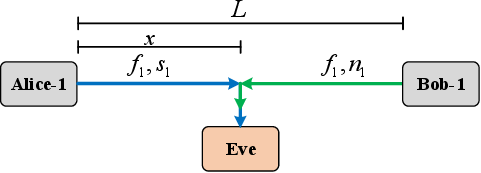}}
	\caption{Illustration of the secure telephone system.}
	\label{fig1}
\end{figure}

As shown in Fig.\;\ref{fig1}, we consider a bidirectional wire-line telephone system consisting of two terminals, Terminal-A and Terminal-B, connected by a telephone line.  Signal transmission from Terminal-A to Terminal-B occurs through Alice-1 to Bob-1 using carrier frequency \(f_1\). Conversely, transmission from Terminal-B to Terminal-A occurs through Alice-2 to Bob-2 using carrier frequency \(f_2\).  The two frequencies, \(f_1\) and \(f_2\), enable separation of the bidirectional transmissions at the receivers by the filters.  The PLS model is in the presence of an eavesdropper, Eve, who attempts to detect the transmitted signals at \(f_1\) and \(f_2\) on the telephone line between the two terminals.  Since the bidirectional transmissions can be treated as two independent problems by using filters, we can simplify the model to the single-directional signal transmission.  Thus, the following discussions will use this simplified model.

\subsection{Security Model}
In this subsection, we focus on the secure communication problem for single-directional transmission, as shown in Fig.\;\ref{fig1b}.  In this setup, Terminal-A transmits the signal \(s_1\) to Terminal-B through Alice-1 and Bob-1.  Meanwhile, the AN, \(n_1\), is generated by Bob-1 to interfere with Eve's reception.  To ensure secure communication, the AN is eliminated at Bob-1.  The power of the signals and ANs are assumed to be $P_s$ and $P_n$, respectively.  Therefore, the signal intercepted by Eve is a mixture of the transmitted signal and the AN.  When Eve is positioned at a distance \(x\) from Alice, the intercepted signal can be expressed as
\begin{align} \label{r_Eve}
	r_{\rm Eve}(x,t) &= \sqrt{\alpha(x)} s_1\left(t-\frac{x}{c}\right) \nonumber\\
	& \hspace{4mm} + \sqrt{\alpha(L-x)} n_1\left(t-\frac{L-x}{c}\right) + n_{\rm Eve}(t),
\end{align}
where $n_{\rm Eve}(t)$ represents the thermal noise at Eve, $c$ is the velocity of light, and $\alpha(x)$ is the attenuation factor that describes the decay in signal power as it travels along the wire-line.

The signal received by Bob-1, after AN cancellation, can be written as 
\begin{equation} \label{r_Bob}
	r_{\rm Bob}(t) = \sqrt{\alpha(L)} s_1\left(t-\frac{L}{c}\right) + \beta n_1(t) + n_{\rm Bob}(t),
\end{equation}
where $\beta$ indicates the AN cancellation capability and $n_{\rm Bob}(t)$ is the thermal noise of Bob.  Both thermal noises above are assumed to have a power of $\sigma_n^2$.  A strong cancellation capability ensures that $\beta \ll 1$.

From \eqref{r_Eve} and \eqref{r_Bob}, it can be observed that, on one hand, we aim for Bob to completely eliminate the AN, i.e., $\beta = 0$, to achieve optimal signal reception.  On the other hand, the power of the AN at Eve's location should be strong enough to obscure the transmitted signal.  This process forms the basis of the proposed PLS model that utilizes the AN to enhance communication security.

\subsection{Secrecy Capacity}
To ensure absolute security, we consider the detection of secret messages across both temporal and spatial dimensions.  The maximum energy of the signals potentially intercepted by Eve depends on integrating over the entire symbol duration \(T\).  Given multiple interception opportunities, Eve can occasionally synchronize with the symbol, which represents the optimal detection scenario.

The transmission delays, \(\frac{x}{c}\) and \(\frac{L-x}{c}\), are assumed to be negligible because they are significantly shorter than the symbol duration.  Without loss of generality, we focus on the first symbol.  After combining all the intercepted signals within this interval, we can express the signals intercepted by Eve as
\begin{align}
	\hat{r}_\mathrm{Eve}^{x} &= \frac{1}{T}\int_0^T {r_\mathrm{Eve}^{x}(t) \mathop{}\!\mathrm{d}t} \nonumber \\
	&= \sqrt{\alpha(x)}s_1(0) + \sqrt{\alpha(L - x)} \hat{n}_1  + \hat{n}_{\rm Eve},
\end{align}
where \(\hat{n}_1 = \frac{1}{T}\int_0^T{n_1(t)}\mathop{}\!\mathrm{d}t\) and \(\hat{n}_{\rm Eve} = \frac{1}{T}\int_0^T{n_{\rm Eve}(t)}\mathop{}\!\mathrm{d}t\). Likewise, the signals Bob receives in the first symbol are expressed as
\begin{align}
	\hat{r}_\mathrm{Bob} &= \frac{1}{T}\int_0^T {r_\mathrm{Bob}(t)\mathop{}\! \mathrm{d}t} \nonumber \\
	&= \sqrt{\alpha(L)}s_1(0) + \sqrt{\beta} \hat{n}_1 + \hat{n}_{\rm Bob},
\end{align}
where \(\hat{n}_1= \frac{1}{T}\int_0^T{n_1(t)}\mathop{}\!\mathrm{d}t\) and \(\hat{n}_{\rm Bob} = \frac{1}{T}\int_0^T{n_{\rm Bob}(t)}\mathop{}\!\mathrm{d}t\).

From the perspective of colluding eavesdropping, Eve may combine the results of multiple interceptions to maximize the signal-to-interference-plus-noise ratio (SINR).  Thus, the SINR of Eve's interception is calculated as
\begin{equation}
	\gamma_\mathrm{Eve}^{x} = \frac{\alpha(x) \mathbb{E}\left[\left|s_a(0)\right|^2\right]}{\alpha(L-x)\mathbb{E}\left[\left|\hat{n}_b\right|^2\right] + \sigma_n^2} 
	= \frac{\alpha(x) P_s}{\alpha(L-x) P_n + \sigma_n^2},
\end{equation}
where \(\mathbb{E}[\cdot]\) denotes the mathematical expectation of a random variable.  Clearly, the worst-case scenario for secure communication occurs when Eve's detection is closest to the signal source and furthest from the noise source, i.e., at \(x = 0\).  Thus, the maximum SINR at Eve is
\begin{equation} \label{gamma_Eve}
	\gamma_\mathrm{Eve}^{\max} = \frac{P_s}{\alpha(L) P_n + \sigma_n^2}.
\end{equation}
Correspondingly, Bob's SINR is given by
\begin{equation} \label{gamma_Bob}
	\gamma_\mathrm{Bob} = \frac{\alpha(L) P_s}{\beta P_n + \sigma_n^2}.
\end{equation}
The secrecy capacity of our proposed wire-line system is then expressed as
\begin{align} \label{Cs}
	C_s = \left[ \log_2\left( 1 + \gamma_\mathrm{Bob} \right) - \log_2\left(1 + \gamma_\mathrm{Eve}^{\max} \right) \right]^+,
\end{align}
where \([x]^+\) indicates that \(x\) is a non-negative value.  The capacity \(C_s\) represents the maximum rate at which secure and reliable information can be transmitted from a sender to a receiver in the presence of a wiretapper.  For absolute security, we require \(\gamma_\mathrm{Bob} > \gamma_\mathrm{Eve}^{\max}\), then leading to the conditions
\begin{equation} \label{C}
	\begin{cases}
		P_n > \frac{1 - \alpha(L)}{\alpha^2(L) - \beta} \sigma_n^2,\\
		\alpha^2(L) - \beta > 0.
	\end{cases}
\end{equation}

As can be seen from \eqref{C}, the AN cancellation capability \(\beta\) can be significantly smaller than the wire-line's power attenuation factor \(\alpha(L)\).  Therefore, the inequality is easily satisfied in practice. It is also important to note that \eqref{C} establishes the necessary conditions for ensuring absolutely secure transmission in the wire-line communication system. Clearly, as the length of the wire-line increases, there is a corresponding need for greater AN cancellation capability.

\section{Hardware Implementation} \label{sec3}
\begin{figure}
	\centering
	\subfigure[Schematic]{
		\centering
		\includegraphics[width=0.95\columnwidth]{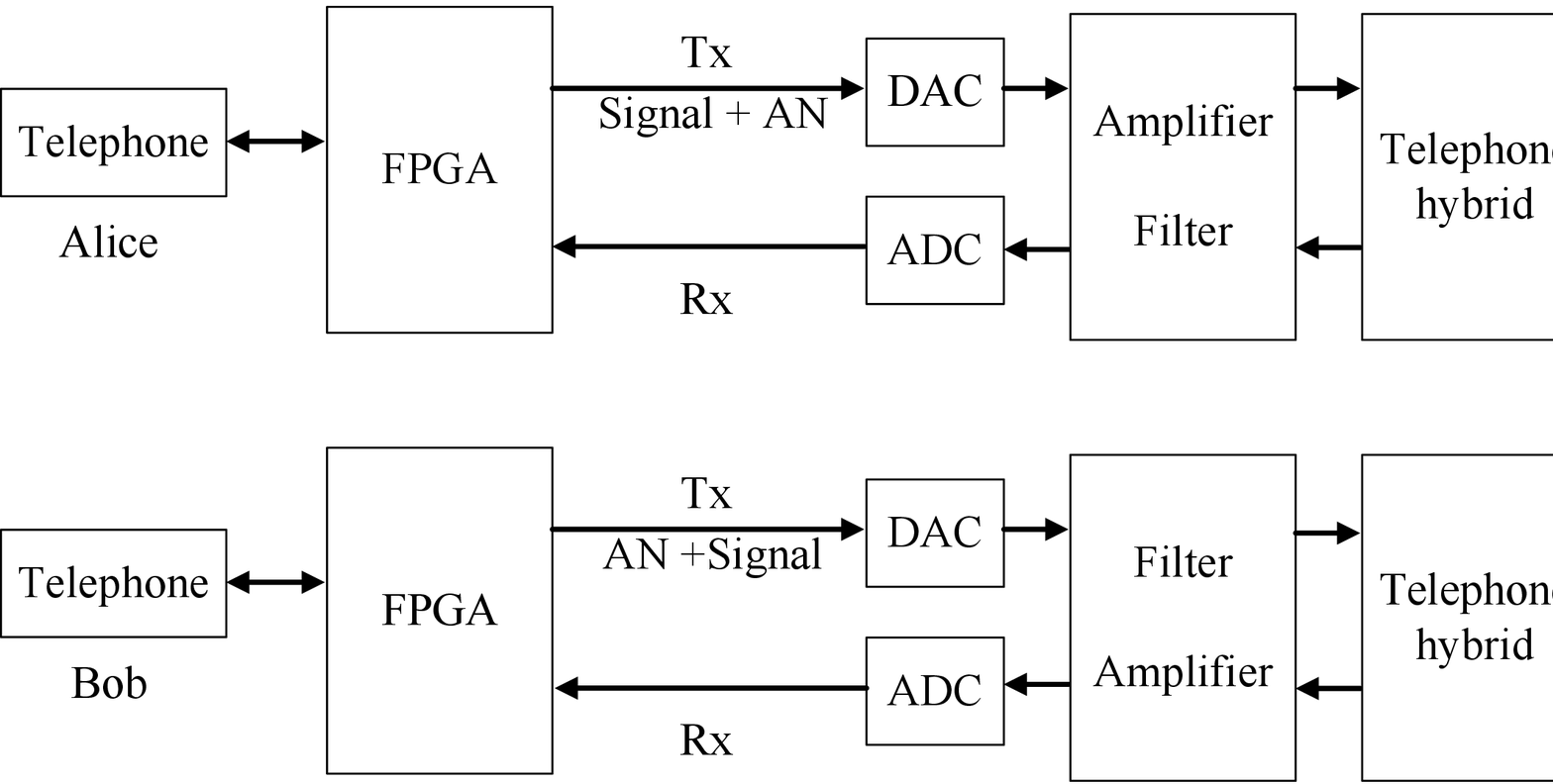} \label{fig2a}
	} \hspace{10mm}
	\subfigure[Implementation]{
		\centering
		\includegraphics[width=0.95\columnwidth]{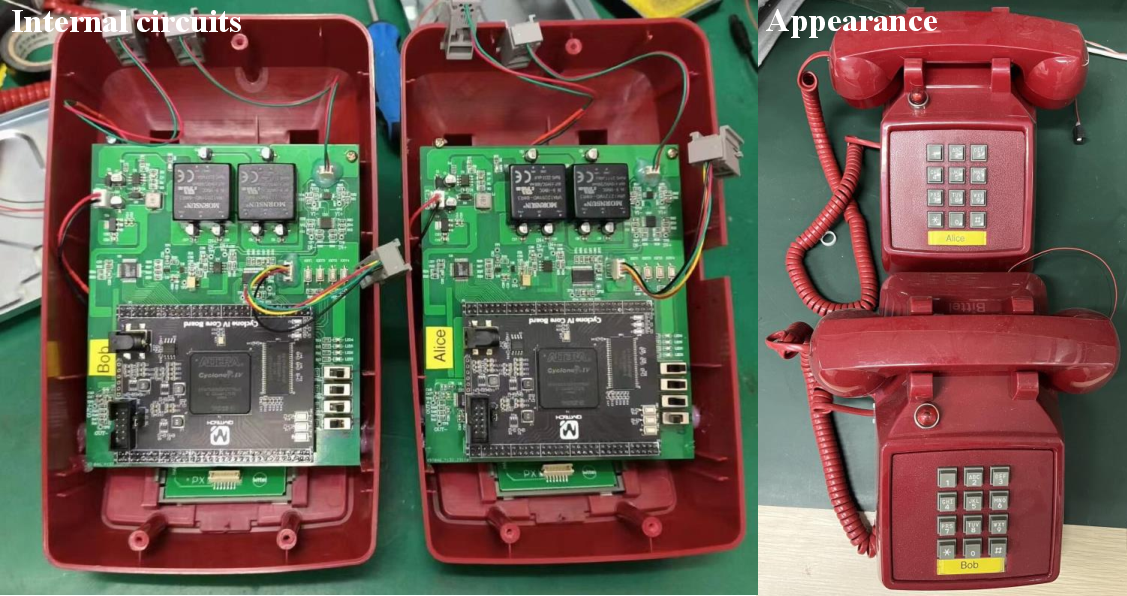} \label{fig2b}
	}
	\caption{Architecture of the secure telephone system.}
	\label{fig2}
\end{figure}

The architecture of the proposed system, consisting of Terminal-A and Terminal-B, is shown in Fig.\;\ref{fig2a}.  Terminal-A transmits signal \(s_1\) from Alice-1 at frequency \(f_1\) to Bob-1 at Terminal-B.  To protect \(s_1\), the AN generator from Terminal-B transmits the AN \(n_1\) at frequency \(f_1\).  Similarly, Terminal-B transmits signal \(s_2\) from Alice-2 at frequency \(f_2\) to Bob-2 at Terminal-A, and the AN generator from Terminal-A transmits the AN \(n_2\) at frequency \(f_2\) to protect \(s_2\).  For the receivers, Bob-1 and Bob-2, the transmitted signals and the AN are eliminated by the telephone hybrids, while the AN from the opposite direction is reduced by the filters.

As shown in Fig.\;\ref{fig2b}, the proposed system is implemented via a field-programmable gate array (FPGA) module (i.e., Intel Cyclone V SE FPGA board), analog-to-digital converters (ADCs, i.e., AD9226), digital-to-analog converters (DACs, i.e., AD9767), amplifiers (i.e., AD8065), and filters (i.e., AD8138).  Furthermore, the proposed system requires a much higher capability for the AN cancellation compared to traditional telephone hybrids, which are designed primarily for echo cancellation.  Therefore, we will explain the circuit design and capability analysis in the following discussions.

\begin{figure}
	\centering
	\subfigure[Schematic]{
		\centering
		\includegraphics[width=0.45\columnwidth]{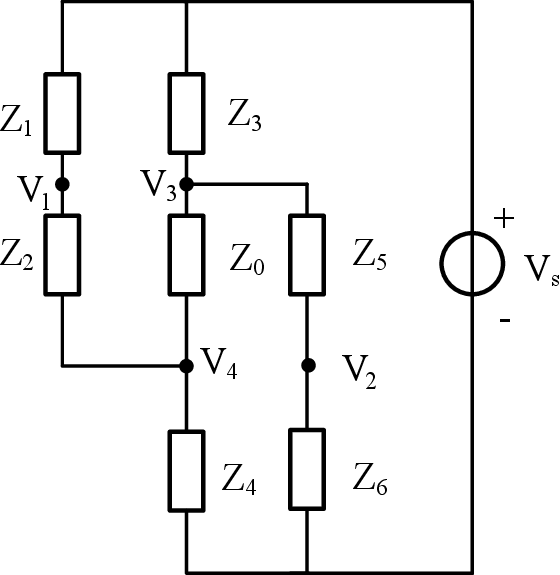} \label{fig3a}
	}
	\subfigure[Mesh current method]{
		\centering
		\includegraphics[width=0.45\columnwidth]{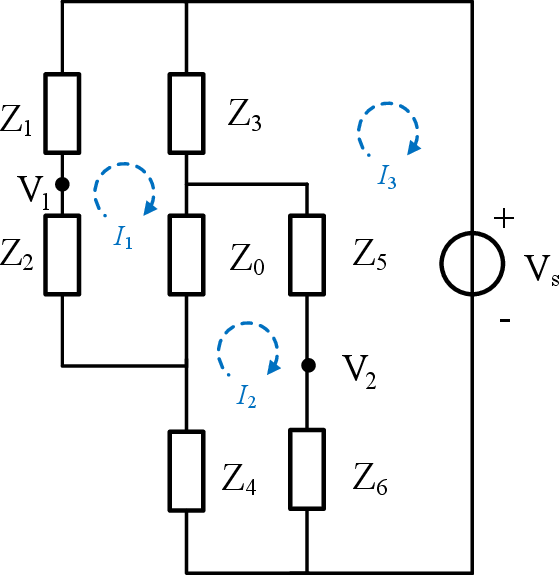} \label{fig3b}
	}
	\caption{Illustration of the telephone hybrid circuit.}
	\label{fig3}
\end{figure}

The telephone hybrid circuit including six resistors \(Z_i,~i=1,2,\cdots,6\) and an equivalent impedance \(Z_0\) is depicted in Fig.\;\ref{fig3}.  From Alice's perspective, Bob is her load; therefore, \(Z_0\) represents the equivalent input impedance of Bob's circuit, and vice versa. \(V_s\) is an equivalent voltage source that generates the transmitted signals. The received signal is denoted by \(V_1 - V_2\), and \(V_3\) and \(V_4\) are the differential voltages on the wire-line. To evaluate the AN cancellation capability, we assume that \(V_s\) represents the transmitted AN and \(V_r = V_1 - V_2\) is the residual AN.

Applying the mesh current method and Kirchhoff's voltage law to each loop, we obtain the following linear equations in matrix form, i.e., \(\mathbf{Z}\mathbf{I} = \mathbf{V}\), with \(\mathbf{I} = \left[I_1,I_2,I_3\right]^\mathrm{T}\), \(\mathbf{V} = \left[ 0,0,-V_s \right]^\mathrm{T}\), and
\begin{equation}
	\mathbf{Z} = \left[ {\begin{array}{*{20}{c}}
			{{{\rm{Z}}_0}{\rm{ + }}{{\rm{Z}}_1}{\rm{ + }}{{\rm{Z}}_2}{\rm{ + }}{{\rm{Z}}_3}}&{{\rm{ - }}{{\rm{Z}}_0}}&{{\rm{ - }}{{\rm{Z}}_3}}\\
			{{\rm{ - }}{{\rm{Z}}_0}}&{{{\rm{Z}}_0}{\rm{ + }}{{\rm{Z}}_4}{\rm{ + }}{{\rm{Z}}_5}{\rm{ + }}{{\rm{Z}}_6}}&{{\rm{ - }}{{\rm{Z}}_5}{\rm{ - }}{{\rm{Z}}_6}}\\
			{{\rm{ - }}{{\rm{Z}}_3}}&{{\rm{ - }}{{\rm{Z}}_5}{\rm{ - }}{{\rm{Z}}_6}}&{{{\rm{Z}}_3}{\rm{ + }}{{\rm{Z}}_5}{\rm{ + }}{{\rm{Z}}_6}}
	\end{array}} \right].
\end{equation}

Under Cramer's rule, the mesh currents can be solved as
\begin{equation} \label{Ii}
	I_i = \frac{\det(\mathbf{Z}_i)}{\det(\mathbf{Z})},~i = 1,2,3,
\end{equation}
where \(\mathbf{Z}_i\) denotes the matrix formed by replacing the \(i\)-th column of \(\mathbf{Z}\) with \(\mathbf{V}\), and \(\det(\cdot)\) denotes the determinant of a matrix.

Moreover, based on Kirchhoff's voltage law, the voltages \(V_1\) and \(V_2\) can be calculated as
\begin{align} \label{V12}
	\left\{ {
		\begin{array}{l}
			V_1 = V_s + I_1Z_1\\
			V_2 = (I_2 - I_3) Z_6
		\end{array}.
	} \right.
\end{align}
With \eqref{Ii} and \eqref{V12}, the residual AN can be expressed as
\begin{equation}
	V_r = V_1 - V_2 = V_s \times Z_\chi,
\end{equation}
where $Z_\chi$ is extremely complicated in its expression and difficult to handle.  However, in practical applications, the residual AN can be further approximated and simplified. Specifically, the resistance values are designed to satisfy \(Z_1,Z_2,Z_5,Z_6\gg Z_3,Z_4\). In this case, the output resistance \(Z_0\) approximates to \(Z_3+Z_4\), i.e., \(Z_0 \approx Z_3+Z_4\). The residual AN can thus be simplified by neglecting the minor terms. As a result, we have
\begin{align}
	V_r \approx V_s&\left[ {\frac{Z_4Z_5-Z_0Z_6}{(Z_3+Z_4+Z_0)(Z_5+Z_6)}} \right. \nonumber \\
	&\hspace{20mm} + \left. {\frac{Z_2Z_3+Z_0Z_2}{(Z_3+Z_4+Z_0)(Z_1+Z_2)}} \right].
\end{align}

\begin{figure}
	\centering
	\includegraphics[width=0.95\linewidth]{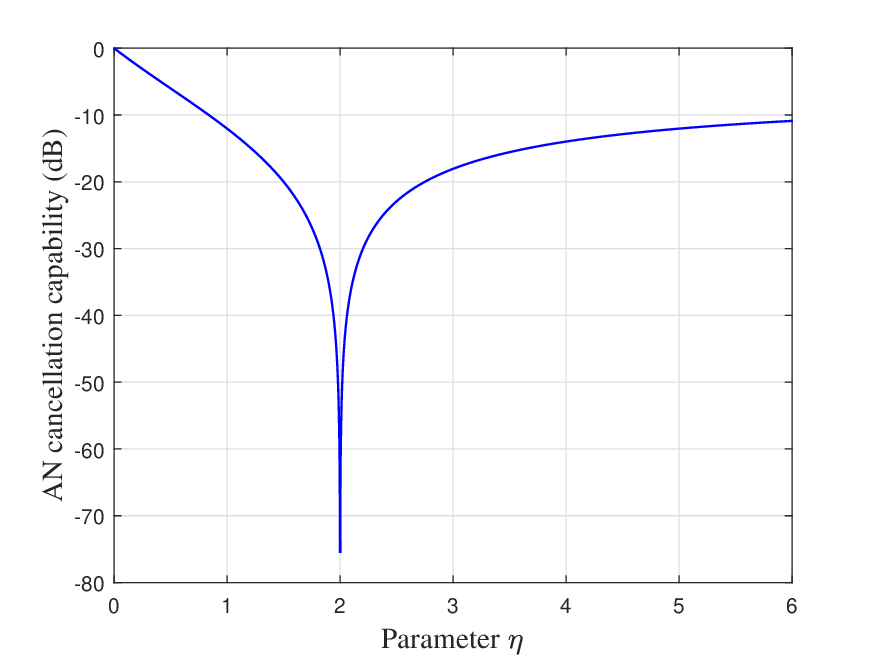}
	\caption{AN cancellation capability versus parameter \(\eta\).}
	\label{fig4}
\end{figure}

In practice, the resistance values are carefully designed with \(Z_1 = Z_6\), \(Z_2 = Z_5\), and \(Z_3 = Z_4\). Recalling that \(Z_1,Z_2,Z_5,Z_6\gg Z_3,Z_4\), we further assume \(Z_3 = Z_4 \buildrel \Delta \over = Z_S\), \(Z_2 = Z_5 \buildrel \Delta \over = Z_L\), and \(Z_1 = Z_6 \buildrel \Delta \over = \eta Z_L\). With these notations, we obtain \(Z_0 \approx Z_3 + Z_4 = 2Z_S\). Finally, the residual AN can be simplified and expressed as
\begin{align} \label{VrVs}
	V_r & \approx V_s\left[\frac{Z_SZ_L-2\eta Z_SZ_L}{4(1+\eta)Z_SZ_L} + \frac{Z_SZ_L+2Z_SZ_L}{4(1+\eta)Z_SZ_L}\right] \nonumber \\
	& = \frac{2-\eta}{2+2\eta}V_s = \beta V_s.
\end{align}
As shown in \eqref{VrVs} and illustrated in Fig.\;\ref{fig4}, for maximum AN cancellation capability, the parameter \(\eta\) should be set to two.

\section{Performance Evaluation} \label{sec4}
In this section, we carry out simulations and experiments to demonstrate the AN cancellation capability and security performance of the proposed system.  We set the signal power \(P_s = 0\) dBm, the AN power \(P_n = 20\) dBm, and the noise power \(\sigma_n^2 = -100\) dBm, with the signal strength attenuating by 2 dB per hundred meters over the wire-line \cite{ref_Liu_CL_2017}.  For the telephone hybrid design, we use resistors with values \(Z_3 = Z_4 = 50\;\Omega\), \(Z_1 = Z_6 = 2\;{\rm K}\Omega\), and \(Z_2 = Z_5 = 1\;{\rm K}\Omega\).

\begin{figure}
	\centering
	\includegraphics[width=0.95\columnwidth]{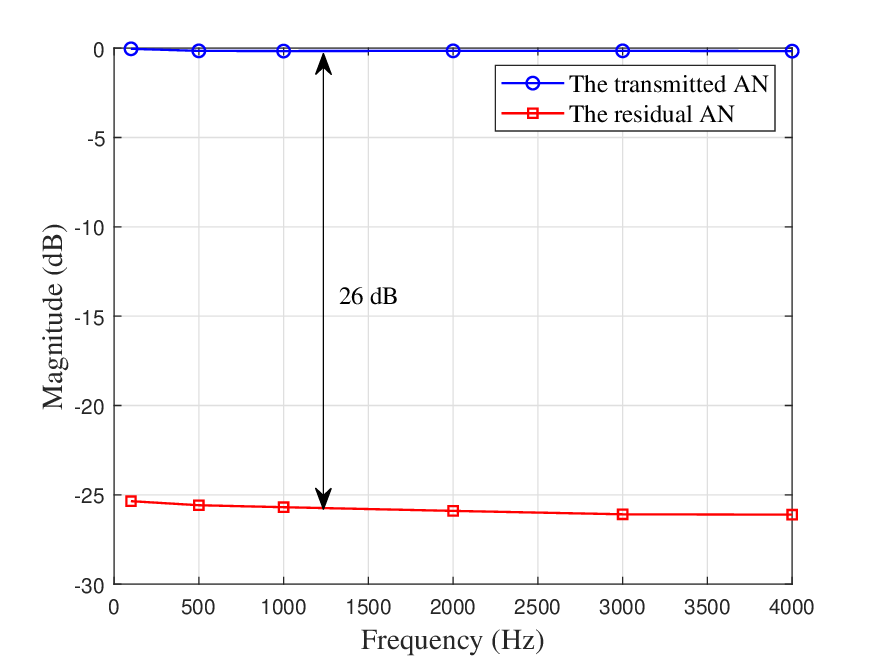}
	\caption{Frequency response of the proposed system.}
	\label{fig5}
\end{figure}

First, the measured frequency response of the proposed system for the generated AN is shown in Fig.\;\ref{fig5}. In these measurements, a single-frequency sinusoid ranging from 0 Hz to 4 kHz, used to mimic the AN, is generated by the AN generator. We captured the transmitted AN at position \(x = 0\) meters and the residual AN after cancellation by the telephone hybrid. Both the transmitted and residual AN are plotted in the figure. The results demonstrate that the hardware prototype achieves significant AN cancellation, with approximately 26 dB of suppression. This high level of cancellation effectively minimizes the AN, which prevents it from interfering with the desired signals. Importantly, the desired signals are preserved with only negligible degradation and it confirms that signal quality is maintained throughout the process.  Furthermore, the frequency response of both the transmitted and residual AN remains remarkably consistent and flat up to 4 kHz.  This stability indicates that the system performs reliably across this frequency range without introducing significant distortions or variations. It also highlights the effectiveness of the hardware design in maintaining signal fidelity while providing robust AN cancellation.

\begin{figure}
	\centering
	\includegraphics[width=0.95\columnwidth]{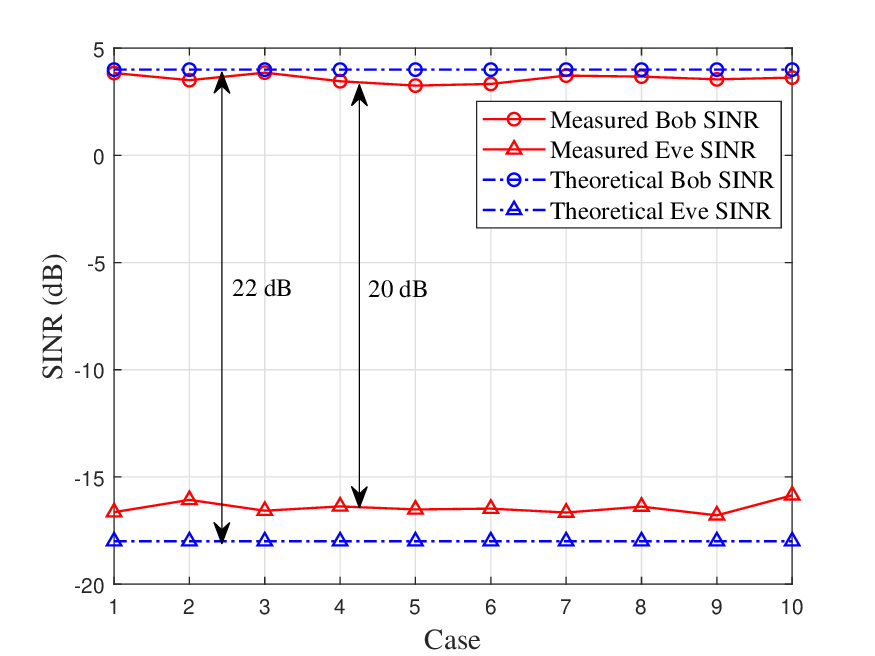}
	\caption{Comparison of SINRs of legitimate and eavesdropping links.}
	\label{fig6}
\end{figure}

Next, the SINRs of the legitimate and eavesdropping links are exhibited in Fig.\;\ref{fig6}.  In this setup, the communication link from Alice-1 to Bob-1 operates at \(f_1 = 500\) Hz, while the reverse link operates at \(f_2 = 2\) kHz.  The test signals, each lasting one second, are superposed with the AN from Terminal-A and the combined signals are then transmitted through a 100-meter wire-line to Terminal-B.  For comparison, the theoretical SINRs are calculated using \eqref{gamma_Eve} and \eqref{gamma_Bob} by assuming \(\beta = 26\) dB (measured in Fig.\;\ref{fig5}).  The results show that, in each case, the SINR of the legitimate link is consistently about 20 dB higher than that of Eve's link, closely matching the theoretical performance of 22 dB.  This consistent SINR difference highlights the system's robust ability to secure the communication channel against eavesdropping.  The slight 2 dB degradation from the theoretical value indicates that the system is functioning near maximum levels, effectively balancing performance and security.  Notably, this 20 dB advantage is observed across multiple cases, thereby underscoring the stability and reliability of the system performance.

\begin{figure}
	\centering
	\includegraphics[width=0.95\columnwidth]{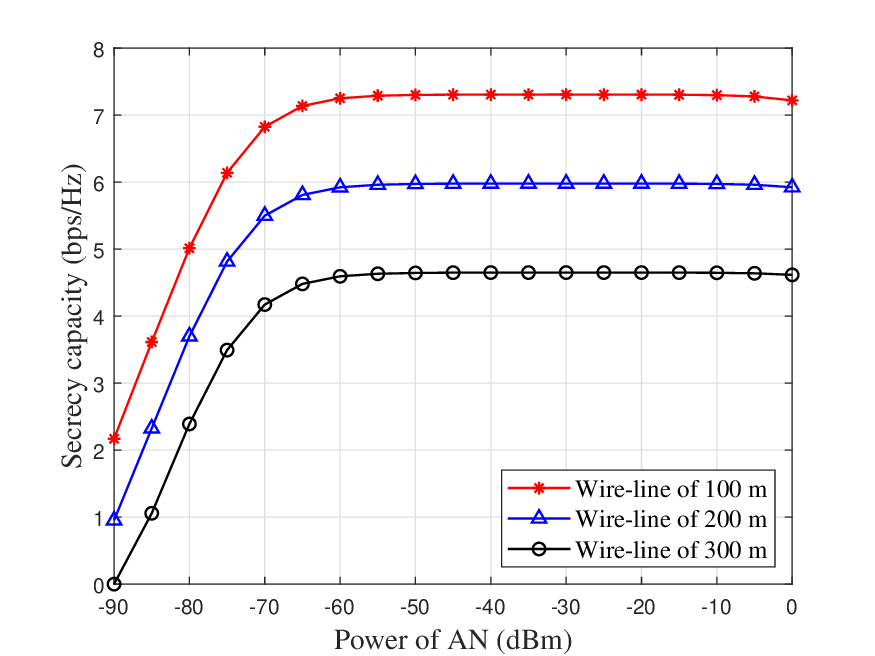}
	\caption{Secrecy capacity versus the power of AN.}
	\label{fig7}
\end{figure}

Lastly, Fig.\;\ref{fig7} illustrates the relationship between secrecy capacity and AN power under a measured AN cancellation capability of 26 dB.  The secrecy capacity is computed using \eqref{Cs} for wire-line lengths of 100, 200, and 300 meters.  From the figure, it is evident that as the wire-line length increases, the secrecy capacity decreases.  This is due to the greater signal attenuation over longer distances, which reduces the SINR advantage for the legitimate receiver and makes it easier for eavesdroppers to intercept the signal.  Moreover, the figure shows that increasing the power of AN significantly enhances the system's secrecy capacity.  This improvement occurs because stronger AN creates more interference for Eve, effectively masking the legitimate signal and making it more difficult for unauthorized parties to extract useful information.  As a result, the system's overall security is bolstered.  These findings emphasize the importance of balancing wire-line length and AN power to optimize secrecy capacity.  In scenarios with longer wire-line distances, increasing AN power becomes particularly crucial to maintaining a high level of security in the communication system.

\section{Conclusion} \label{sec5}
In this paper, we proposed a secure wire-line telephone system that uses a telephone hybrid circuit to effectively eliminate the AN.  We presented the design principles and derived the AN cancellation capability for specific resistance values based on differential signals.  Our analysis showed that this capability significantly affects the performance of legitimate communication links.  The system performance was validated through simulations and experiments, and the results demonstrated its effectiveness in maintaining high secrecy capacity under various conditions.

\end{document}